\documentclass[twocolumn,showpacs,preprintnumbers,amsmath,amssymb]{revtex4-1}

\usepackage{epsfig}
\usepackage{graphicx}
\usepackage{dcolumn}
\usepackage{bm} 
\usepackage{subfigure}
\usepackage{appendix}

\newcommand{\beq}{\begin{equation}}
\newcommand{\eeq}{\end{equation}}
\newcommand{\beqa}{\begin{eqnarray}}
\newcommand{\eeqa}{\end{eqnarray}}

\newcommand{\la}{\langle}
\newcommand{\ra}{\rangle}
\newcommand{\ket}[1]{|#1\rangle}                
\newcommand{\bra}[1]{\langle#1|}                
\newcommand{\scpr}[2]{\langle#1|#2\rangle}      
\newcommand{\matel}[3]{\langle#1|#2|#3\rangle}  
\newcommand{\Tr}[1]{\mbox{Tr}_{#1}}              
\DeclareRobustCommand{\orderof}{\ensuremath{\mathcal{O}}}

\def\pra#1{{ Phys.\ Rev. A\/} {\bf#1}}
\def\prb#1{{ Phys.\ Rev. B\/} {\bf#1}}

\def\prl#1{{ Phys.\ Rev.\ Lett.} {\bf#1}}

\begin{document}

\title{Dissipation of the Rabi model beyond the rotating wave approximation: Quasi-degenerate qubit and ultra-strong coupling}

\author{S. Agarwal, S.M. Hashemi Rafsanjani and J.H. Eberly}

\affiliation{ Rochester Theory Center and the Department of Physics \& Astronomy\\
University of Rochester, Rochester, New York 14627}

\email{shantanu@pas.rochester.edu}


\date{\today}

\begin{abstract}
Environmental influences on the dynamics of a coupled qubit-oscillator system are studied analytically. We investigate the case of a quasi-degenerate qubit within the ultra-strong coupling regime for which the qubit frequency is much smaller than the frequency of the oscillator, and the coupling between the qubit and the oscillator is large, both of which invalidate the usually employed rotating wave approximation. In contrast to the standard quantum optics master equation, we explicitly take the qubit-oscillator coupling into account while microscopically deriving a dressed state master equation. Using the derived master equation, we discuss a spectroscopic technique which can be used to probe the dressed energy level structure of the qubit-oscillator system.
\end{abstract}



\maketitle


\section{Introduction}


A two-level system (qubit) that interacts with a harmonic oscillator has been the cornerstone of various experimental and theoretical studies \cite{Walther-06, Blais-etal}. In many of these physical scenarios, the dynamics of the qubit-oscillator system is governed by the Rabi Hamiltonian \cite{Rabi}:
\begin{align} \label{e.Ham_Rabi} 
H_{Rabi}=\hbar\omega_{0}\sigma_{z}/2 + \hbar\omega a^{\dagger}a+ \hbar\beta\omega(a+a^{\dagger})\sigma_{x}.
\end{align}

In the historically most common experimental context of cavity QED, the coupling is extremely small (i.e., $\beta \lesssim 10^{-6}$) and the qubit and the oscillator are nearly resonant (i.e. $\omega_0\sim\omega$) \cite{Walther-06}. Under these conditions, it is valid to make an approximation of dropping the counter terms, $a\sigma_-$ and $a^\dagger\sigma_+$ where $\sigma_x=(\sigma_++\sigma_-)$, from the interaction Hamiltonian. This is the well-known Jaynes-Cummings domain of the rotating wave approximation (RWA), where remarkably simple exact solutions are available for eigen-energies and eigenvectors of $H_{Rabi}$ \cite{JC}. The effect of the counter rotating terms become important either because of ultra-strong coupling, $|\beta|\gtrsim0.1$, or because of extremely large detuning, $|\omega-\omega_0|\sim\omega+\omega_0$. Under these conditions, the RWA cannot be made \cite{Allen-Eberly}. 

Apart from the coherent dynamics generated by $H_{Rabi}$, there is incoherent evolution due to the system's unavoidable interaction with its environment. The dissipative dynamics of the qubit-oscillator system has been investigated extensively, focusing mostly on the regime where the RWA is valid \cite{Agarwal-86, Barnett-86, Eislet-91, Werner-91, Piexoto-99, Boissonneault-08}. The incoherent evolution in this regime is phenomenologically modeled by the standard quantum optics master equation (SME) which assumes that the qubit-oscillator coupling plays no role in the dissipation mechanism. This approximation leads the SME to have non-physical predictions \cite{Carmichael-73, Cresser-92, Zoubi-00,Scala-07,Beaudoin-11}. Although the predictions of the SME are not strictly correct, the errors in its predictions are small when the coupling is small. However, the SME must be abandoned when the coupling is ultra-strong to invalidate the RWA. 

Developments in the area of circuit QED have made it possible to engineer systems that operate in regimes where the RWA breaks down \cite{Niemczyk-etal, Forn-etal, Fedorov-etal}. Thus, the question of damping in the regime beyond the RWA has gained practical importance and has recently begun to be theoretically explored \cite{Beaudoin-11, Hausinger-08, Liberato-09, Cao-11, Ridolfo-12}. All these investigations focus on understanding the effects of increasing the coupling strength on the damped dynamics while keeping the qubit nearly resonant with the oscillator. 

Damping in the ultra-strong but far off-resonance regime has not yet been analyzed in the literature. The aim of this report is to explore this parameter space. In particular, we investigate the quasi-degenerate qubit regime where the qubit is far red-detuned from the oscillator ($\omega_0\ll\omega$) and the coupling strength is ultra-strong \cite{Casanova-10}.  As the SME becomes invalid in this parameter domain, we adopt an alternate approach and follow Carmichael's treatment \cite{Carmichael-73} of investigating the dissipative dynamics of strongly coupled systems by microscopically deriving the composite system's dressed state master equation (DME). The DME approach has previously been used to study the dissipative dynamics of ultra-strongly coupled qubit-oscillator systems while restricting attention only to the near resonance regime \cite{Beaudoin-11, Hausinger-08, Ridolfo-12}. Here we use the DME approach to explore the dissipative dynamics of ultra-strongly coupled qubit-oscillator system in the regime where the qubit is quasi-degenarate and is thus far away from resonance with the oscillator. 

In contrast to the SME, it will be shown that the DME predicts the correct steady state of the interacting qubit-oscillator system at thermal equilibrium. In the derivation of the DME, we model the qubit and the oscillator environments by separate bosonic baths. The effect of the energy level fluctuations of the qubit is also considered. Using the DME, we suggest a spectroscopic technique which can be used to probe the dressed energy level structure of the Rabi Hamiltonian in the quasi-degenerate qubit and ultra-strong coupling regime. 


\section{Adiabatic approximation}\label{s.Ad-Approx}


In order to derive the dressed state master equation, we need to find the eigen-energies and eigenstates of the Rabi Hamiltonian in the parameter domain of interest where $\omega_0 \ll \omega$ and the coupling is ultra-strong and thus the RWA becomes invalid. For this purpose, we use a form of an \textit{adiabatic approximation} which was previously presented by Irish, et al. \cite{Irish-05}. Here, we quickly sketch this approximation. 

Because $\omega_0$ is so small, the terms of zero-order and first-order importance in $H_{Rabi}$ are completely different from those in the JC Hamiltonian. Here $\frac{1}{2}\hbar\omega_0\sigma_z$ can be ignored in zero-order, and the remainder of the Hamiltonian provides the eigen-energies and eigenstates forming the basis for the dynamical evolution.   

When ignoring the $\sigma_z$ term, the $\sigma_x$ operator is constant, so we work in its eigenbasis and use $m = \pm 1$ to designate its eigenvalues. In its eigenbasis the zero-order Hamiltonian can be quickly diagonalized by factoring:
\beqa \label{H_AD} 
H_0 &=& \hbar\omega(a^\dag + m\beta)(a + m\beta) - \hbar\omega\beta^2, \nonumber\\
&=& \hbar\omega b^\dag b - \hbar\omega\beta^2, 
\eeqa
where the new $b$ operators are obviously displaced versions of the $a$ operators, and we have
\beq 
H_0 \ket{m,N_m} = \mathcal{E}_N  \ket{m,N_m},
\eeq
where
\beq\label{e.En}
\mathcal{E}_N=\hbar\omega(N-\beta^2).
\eeq
The eigenfunctions of $H_0$ clearly have the form 
$$\ket{m,N_m}=\ket{m}\otimes\ket{N_m},$$ 
where $\ket{m=\pm}$ are the eigenstates of $\sigma_x$ and $\ket{N_m}$ are oscillator states defined by:
\begin{align}\label{e.dis_Fock}
\ket{N_m}&=\exp\Big(-m\beta(a^{\dagger}-a)\Big)\ket{N},\nonumber\\
&=D(-m\beta)\ket{N},
\end{align}       
where $D(m\beta)$ is a displacement operator. The states, $\ket{N}$, are number states of the oscillator: $a^{\dagger}a\ket{N}=N\ket{N}$. We see from Eq. (\ref{e.dis_Fock}) that $\ket{N_m}$ are displaced Fock states, the displacement of which depends upon the state of the qubit. The energy eigenstates corresponding to the same $N$: $\ket{-,N_-}$ and $\ket{+,N_+}$, are degenerate.

The $\sigma_z$ term in the Rabi Hamiltonian remains to be included. It promotes transitions among the $H_0$ eigenstates, and it has non-zero matrix elements that need to be organized systematically. A straightforward approach is to write out the Rabi Hamiltonian matrix in the $H_0$ eigenbasis, including all of the $\sigma_z$ contributions, in order to show that almost all of them can be ignored on the basis of an adiabatic approximation that can be seen as a natural extension of the rotating wave approximation \cite{Irish-05}. That is, fast terms are identified and eliminated in favor of slowly varying quasi-static terms that are retained. The full $H_{Rabi}$ matrix, in stylized form, is displayed here. Rows and columns of $H_0$ are ordered in the sequence $\ket{-,1_{-}},\ket{+,1_{+}},\ket{-,2_{-}},\ket{+,2_{+}},\dots$, etc.:

\beq\label{e.H_N}
 H_{Rabi}=
\begin{pmatrix} \mathcal{E}_0 & 00 & - & 01~ & - & 02~ & \dots \\ 
                        00 & \mathcal{E}_0 & 01 & - & 02 & - & \dots \\
                        - & 10 & \mathcal{E}_1 & 11 & - & 12 & \dots \\
                        10 & - & 11 & \mathcal{E}_1 & 12 & - & \dots \\
                        - & 20 & - & 21 & \mathcal{E}_2 & 22 & \dots \\
                        20 & - & 21 & - & 22 & \mathcal{E}_2 & \dots \\
                        \vdots & \vdots & \vdots & \vdots & \vdots & \vdots & \ddots \\
 \end{pmatrix}.
\eeq
The diagonal elements are the bare energies given in (\ref{e.En}). As shown, there are two $\mathcal{E}_N$ values for each $N$. The dashes in the matrix stand for elements that are zero, and the off-diagonal paired numbers are a simple shorthand for the matrix elements of the $\sigma_z$ term in the Rabi Hamiltonian. That is, 
\beq \label{off-diag element}
MN \equiv \frac{1}{2}\hbar\omega_0 \la \mp,M_{\mp}|\sigma_z|\pm,N_{\pm}\ra.
\eeq

An adiabatic approximation is easily made by noting that every off-diagonal element in $H_{Rabi}$ is associated with an oscillation at the frequency $(M-N)\omega$. Only the diagonal elements and the off-diagonal terms labelled $NN$ are static, i.e., are associated with a zero-frequency oscillation. In our adiabatic approximation, only these terms are kept, implicitly averaging all of the others to zero over a few periods $2\pi/\omega$. That is, we discard all terms with time dependences that are assumed, in this adiabatic approach, to be too rapid to be perceptible - terms oscillating at the (high) oscillator frequency $\omega$ or any of its harmonics. 

The consequence of this adiabatic approximation is a reduced Hamiltonian in block-diagonal form, where the general $N$th block is $2\times2$ dimensional. Each of these blocks is spanned by the states $\ket{-,N_{-}}$ and $\ket{+,N_{+}}$, and has the form:
\beq\label{e.H_N}
 H^{(N)}_{AD}=
\begin{pmatrix} \mathcal{E}_N & NN \\ 
                NN & \mathcal{E}_N\\
 \end{pmatrix},
\eeq
where the off-diagonal elements can be evaluated exactly to \cite{Irish-05}:
\begin{align}\label{e.Rabi_Lag}
NN &\equiv\hbar\frac{\omega_0}{2}\scpr{N_-}{N_+},\nonumber\\
&=\hbar\frac{\omega_0}{2}e^{-2\beta^2}L_{N}(4\beta^2),
\end{align}
with $L_{N}(x)$ being Laguerre polynomials. The eigenstates and the corresponding eigenvalues of $H_N$ are:
\begin{align}\label{e.adiabatic_eigen}
\ket{\Psi_{N}^{\pm}}&=\frac{1}{\sqrt{2}}\Big(\ket{+,N_+}\pm\ket{-,N_-}\Big),\nonumber\\
E_{N}^{\pm}&=\mathcal{E}_{N}\pm\hbar\frac{\omega_0}{2}\scpr{N_-}{N_+}.
\end{align}
Earlier explanations and uses of the adiabatic approximation can be found in \cite{Irish-05, Hausinger-10,Shantanu-12}

It should be noted that in determining the above eigenvalues and eigen-functions, only the qubit frequency was assumed to be much smaller than the oscillator frequency and no approximation on the strength of the coupling, $|\beta|$, was made. In this report, we do not explore the entire parameter regime spanned by arbitrarily high coupling strengths but restrict our analysis to the regime where terms that are fourth or higher order in the coupling strength can be neglected, $\mathcal{O}(\beta^4)\sim0$. This focuses our analysis to the parameter domain which is experimentally feasible with current technology or is likely to be realizable within the near future  \cite{Niemczyk-etal, Forn-etal}. In this regime, one can approximate the exponential appearing in Eq. (\ref{e.adiabatic_eigen}) to be:
\begin{align}\label{e.exp_approx}
e^{-2\beta^2}\approx1-2\beta^2.
\end{align}
A modest upper limit on the coupling strength that satisfies the above approximation is $|\beta_{max}|=0.2$. Coupling strengths greater than $|\beta|=0.1$ are considered to be ultra-strong \cite{Casanova-10} as the validity of the RWA is known to break down under such strong interactions. Thus, the approximation, $\mathcal{O}(\beta^4)\sim0$, does not prevent us from exploring the ultra-strong parameter regime although it does restrict us from exploring the deep strong coupling regime where $|\beta|\gtrsim1$ \cite{Casanova-10}.  

If we further restrict our analysis to the parameter regime where the oscillator excitation number obeys $N \ll  N_{max} = 1/(2\beta)^2$, the Laguerre polynomial appearing in Eq. (\ref{e.Rabi_Lag}) can be approximated to:
\begin{align}\label{e.lag_approx}
L_{N}(4\beta^2)\approx1-4N\beta^2.
\end{align} 
Under the approximations given in Eqs. (\ref{e.exp_approx}) and (\ref{e.lag_approx}), the energy eigenvalues in Eq. (\ref{e.adiabatic_eigen}) take the form:
\begin{align}\label{e.en_adiab}
E_{N}^{\pm}&=\mathcal{E}_{N}\pm\hbar\frac{\omega_0}{2}\Big(1-2\beta^2-4N\beta^2\Big),\nonumber\\
&=\hbar N\omega_\pm\pm\hbar\frac{\omega_0}{2}(1-2\beta^2),
\end{align}
where 
\begin{align}\label{e.om_pm}
\omega_\pm=\omega\mp2\omega_0\beta^2.
\end{align}
In the above equation, we have neglected the constant energy term, $-\hbar\omega\beta^2$, appearing in $\mathcal{E}_N$ (see Eq. (\ref{e.En})). 

Note that $E_N^+-E_N^-=\hbar\omega_0(1-2\beta^2-4N\beta^2)$. Since the excitation number is restricted by $N\ll1/(2\beta)^2$, for the parameter regime that we consider we always have $E_N^+>E_N^-$.

We see from Eq. (\ref{e.en_adiab}) that for various values of $N$, the set of energies: $\{E_{N}^+\}$, are equally spaced. This suggests that the corresponding set of eigenstates: $\{\ket{\Psi_{N}^+}\}$, form a harmonic oscillator ladder with frequency $\omega_+$ and with ground state energy $\hbar\omega_0(1-2\beta^2)/2$. We will refer to this oscillator as $\mathcal{HO}^+$. Similarly, the set of eigenstates: $\{\ket{\Psi_{N}^-}\}$, also form a harmonic oscillator ladder with frequency $\omega_-$ and ground state energy $-\hbar\omega_0(1-2\beta^2)/2$. We will refer to this oscillator as $\mathcal{HO}^-$.  

In order to conveniently treat the two harmonic oscillator structures, we define the annihilation operators, $a_{\pm}$, and projection operators, $1_\pm$:
\begin{align}\label{e.apm}
a_\pm&=\sum_{N\ll N_{max}}\sqrt{N+1}\ket{\Psi_{N}^{\pm}}\bra{\Psi_{N+1}^{\pm}},\nonumber\\
1_{\pm}&=\sum_{N\ll N_{max}}\ket{\Psi_{N}^{\pm}}\bra{\Psi_{N}^{\pm}}.
\end{align} 
In terms of these operators and within the adiabatic approximation, the qubit-oscillator Hamiltonian becomes:
\begin{align}\label{e.H_AD}
H_{AD}=&\hbar\Big(\omega_+a^{\dagger}_+a_+ +\frac{\omega_0}{2}(1-2\beta^2)1_+\Big)\nonumber\\
&+\hbar\Big(\omega_-a^{\dagger}_-a_- -\frac{\omega_0}{2}(1-2\beta^2)1_-\Big).
\end{align} 
From the above form of the Hamiltonian within the adiabatic approximation, $H_{AD}$, it is clear that within the parameter regime for which $\omega_0\ll\omega$ and $\mathcal{O}(\beta^4)\sim0$, the qubit-oscillator composite system can be described by a set of two oscillators: $\mathcal{HO}^{+}$ with frequency $\omega_+$ and $\mathcal{HO}^{-}$ with frequency $\omega_-$.    

We reiterate the important point that the validity of the adiabatic approximation relies upon the slowness of the qubit, $\omega_0\ll\omega$. It imposes no restriction on the strength of the coupling parameter $\beta$. However, the choice of analyzing only the parameter space for which $\mathcal{O}(\beta^4)\sim0$, restricts our analysis to the regime where instead of using the adiabatic approximation, one could use the Schrieffer-Wolff (SW) transformation \cite{Zueco} to find the eigen-structure of $H_{Rabi}$. A comparison between the SW transformation and the adiabatic approximation is given in Appendix \ref{a.Comparison}. For the quasi-degenerate qubit regime, it is shown that the adiabatic approximation works better than the SW transformation.


\section{Dissipation}\label{s.Damping}


The dissipative dynamics of the qubit-oscillator system is usually studied using a standard quantum optics master equation (SME) \cite{Agarwal-86,Barnett-86,Eislet-91,Werner-91,Piexoto-99}:
\begin{align}\label{e.SME}
\dot{\rho}=&\frac{1}{i\hbar}\left[ H,\rho \right]+\kappa \mathcal{D}\left[a\right]\rho+\gamma \mathcal{D}\left[\sigma_-\right]\rho,
\end{align}
where $\rho$ is the density matrix of the qubit-oscillator system and $\kappa$ and $\gamma$ are the relaxation rates of the oscillator and the qubit respectively that are coupled to separate zero temperature baths. The dissipator, $\mathcal{D}$, defined as
\begin{align}\label{e.Dissipator}
\mathcal{D}\left[O\right]\rho=\Big(2O\rho O^{\dagger}-O^{\dagger}O\rho-\rho O^{\dagger}O\Big)/2,
\end{align}
generates non-unitary evolution of the system and the effect of the environment on the system dynamics are encoded in it. In the SME, the qubit-oscillator coupling parameter, $\beta$, plays no role in the form of the dissipators and is introduced phenomenologically in the Hamiltonian only to generate unitary evolution \cite{Scala-07}. Because of neglecting the coupling, the SME predicts phenomena that are physically incorrect \cite{Carmichael-73, Cresser-92, Zoubi-00,Scala-07,Beaudoin-11}. For example, if the qubit and the oscillator environments are both at zero temperature, the Boltzmann distribution demands the steady state of the qubit-oscillator system be the ground state of the joint Hamiltonian, $H_{Rabi}$. In contradiction to this basic principle of thermodynamics, the SME incorrectly predicts that the system will evolve out of its ground state even when the baths are at zero temperature. 

In order to avoid the unphysical predictions of the SME, it is necessary to take the qubit-oscillator coupling into account while describing the damped dynamics. This is achieved by deriving the dressed state master equation (DME) of the interacting systems \cite{Carmichael-73}. 

In this section, we study the damping of the qubit-oscillator system operating in the quasi-degenerate qubit and ultra-strong coupling regime where the RWA breaks down. We provide a microscopic derivation of the dressed state master equation with the dressed states being given in Eq. (\ref{e.adiabatic_eigen}). Note that these dressed states were derived using the adiabatic approximation and retaining the counter-rotating terms in the Rabi Hamiltonian. 

We model the reservoirs of the qubit and the oscillator by separate collections of oscillators with free Hamiltonians:
\begin{align}
H_{osc}^{B}=\hbar\sum_{\lambda}\nu_{\lambda}b^{\dagger}_{\lambda}b_{\lambda},\quad H_{qubit}^{B}=\hbar\sum_{\lambda}\mu_{\lambda}c^{\dagger}_{\lambda}c_{\lambda}.
\end{align}
The Hamiltonian governing the interaction of the qubit and the oscillator with their respective reservoirs is taken to be:
\begin{align}\label{e.H_int}
H_{osc}^{I}&=\hbar\sum_{\lambda}h_{\lambda} (b^{\dagger}_{\lambda}+b_{\lambda})(a^{\dagger}+a),\nonumber\\
H_{qubit}^{I}&=\hbar\sum_{\lambda}q_{\lambda} (c^{\dagger}_{\lambda}+c_{\lambda})\sigma_x.
\end{align} 
The coupling strengths of the reservoirs with the oscillator and the qubit are parameterized by $h_\lambda$ and $q_\lambda$ respectively. These coupling strengths are assumed to be small enough to allow us to make the rotating wave approximation and thus throw away the counter terms in the system-reservoir interaction Hamiltonian (see Eq. (\ref{a.e.BS})). It also allows us to make the secular approximation and Born-Markov approximation when deriving the dressed state master equation in Appendix \ref{a.Master_equation}.

When the reservoir is at zero temperature, the DME takes the form:
\begin{align}\label{e.master_zero}
\dot{\rho}(t)&=\frac{1}{i\hbar}\left[ H_{AD},\rho(t) \right]+\mathcal{J}_{osc}\rho(t)+\mathcal{J}_{qbit}\rho(t),
\end{align}
where 
\begin{align}\label{e.Liov}
\mathcal{J}_{osc}\rho(t)=&\Gamma(\omega_+)\mathcal{D}\left[a_+\right]\rho(t)\nonumber\\
&+\Gamma(\omega_-)\mathcal{D}\left[a_-\right]\rho(t)\nonumber\\
&+4\beta^2\sum_{N}\Gamma(\tilde{\omega}_N)\mathcal{D}\left[\ket{\Psi_{N}^{-}}\bra{\Psi_{N}^{+}}\right]\rho(t),\nonumber\\
\mathcal{J}_{qbit}\rho(t)=&\sum_{N}\gamma(\tilde{\omega}_N)\mathcal{D}\left[\ket{\Psi_{N}^{-}}\bra{\Psi_{N}^{+}}\right]\rho(t).
\end{align}
In Eq. (\ref{e.master_zero}), $\rho(t)$ is the qubit-oscillator composite density matrix and $H_{AD}$ is the Hamiltonian derived within the adiabatic approximation in Eq. (\ref{e.H_AD}). The Liouvillians, $\mathcal{J}_{osc}$ and $\mathcal{J}_{qbit}$, in Eq. (\ref{e.Liov}) are super operators that encode the effects of the reservoirs coupled to the oscillator and the qubit respectively. The damping rates associated with these reservoirs are $\Gamma(\omega)$ and $\gamma(\omega)$. The expressions for these damping rates are given in Eqs. (\ref{a.e.Gamma}) and (\ref{a.e.gamma}).

\begin{figure}[ht]
\centering
\includegraphics[width=7cm, height=8cm]{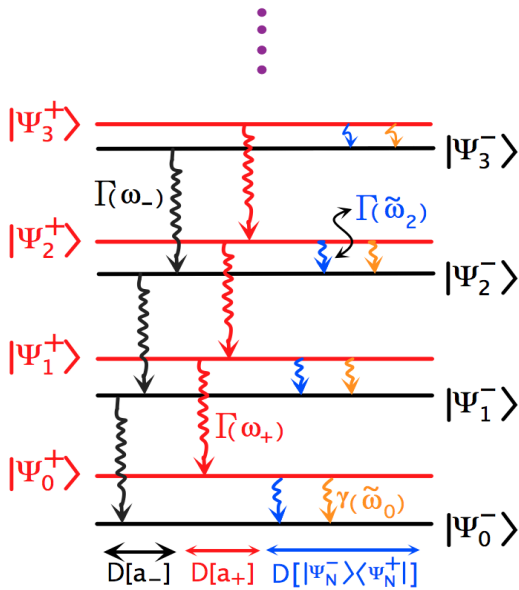}
\caption[Fidelity]{Energy level diagram for the eigenstates, $\ket{\Psi^{\pm}_N}$, of the qubit-oscillator system in the quasi-degenerate qubit and ultra-strong coupling regime where the RWA breaks down. The dissipators, $\mathcal{D}[a_-]$ and $\mathcal{D}[a_+]$, induce transitions within the oscillators $\mathcal{HO^-}$ and $\mathcal{HO^+}$ respectively. Transitions from $\mathcal{HO^+}$ to $\mathcal{HO^-}$ is induced by the dissipator $\mathcal{D}\left[\ket{\Psi_{N}^{-}}\bra{\Psi_{N}^{+}}\right]$. In the long time limit, the steady state of the system is the ground state of the interacting qubit-oscillator system: $\ket{\Psi_0^-}$.}\label{fig:LevelDiag}
\end{figure}

We now try to understand the dynamical implications of the master equation (\ref{e.master_zero}). For doing so, we recall that the energy eigenstructure of the coupled qubit-oscillator system in this regime consists of two harmonic oscillators, $\mathcal{HO}^{\pm}$, with respective frequencies: $\omega_{\pm}$. The energy eigenstates of each of these oscillators are dressed states in the sense that they are entangled qubit-oscillator states. Looking at Eq. (\ref{e.Liov}), we see that the super operators, $\mathcal{J}_{osc}$ and $\mathcal{J}_{qbit}$, consist of dissipators with three different types of lowering operators: $a_+$, $a_-$ and the set of operators $\{\ket{\Psi_{N}^{-}}\bra{\Psi_{N}^{+}}\}$. The term corresponding to the dissipator $\mathcal{D}\left[a_+\right]$, induces damping within the oscillator $\mathcal{HO}^{+}$. Similarly, the term corresponding to the dissipator $\mathcal{D}\left[a_-\right]$, induces damping within the oscillator $\mathcal{HO}^{-}$. The respective damping rates of these oscillators are $\Gamma(\omega_\pm)$. The two oscillators are incoherently coupled by the term corresponding the the dissipator $\mathcal{D}\left[\ket{\Psi_{N}^{-}}\bra{\Psi_{N}^{+}}\right]$ which induces transitions from $\ket{\Psi_N^+}$ to $\ket{\Psi_N^-}$. The term $\mathcal{D}\left[\ket{\Psi_{N}^{-}}\bra{\Psi_{N}^{+}}\right]$ gets contribution from the reservoir connected to the oscillator and the qubit. Corresponding to the reservoir connected to the oscillator and the qubit, the damping rates for the transition $\ket{\Psi_{N}^+}\longrightarrow\ket{\Psi_{N}^-}$ are $\Gamma(\tilde{\omega}_N)$ and $\gamma(\tilde{\omega}_N)$ respectively. These features are schematically shown in Fig. \ref{fig:LevelDiag}.

When a system interacts with zero temperature baths, we expect on thermodynamic considerations that the system should equilibrate to its ground state. Thus, in the present scenario, we expect the steady state of the coupled qubit-oscillator system to be the minimum energy state $\ket{\Psi_0^-}$. To find the steady state predicted by the DME, we have to find the state which does not evolve in time. Putting $\rho(t)=\ket{\Psi_0^-}\bra{\Psi_0^-}$ in Eq. (\ref{e.master_zero}), we easily get that $\dot\rho(t)=0$. This implies that the steady state is $\ket{\Psi_0^-}$ and confirms our earlier statement that at thermal equilibrium, the DME correctly predicts the steady state to be the ground state of the coupled qubit-oscillator system. On the other hand, by putting $\rho(t)=\ket{\Psi_0^-}\bra{\Psi_0^-}$ in the SME given in Eq. (\ref{e.SME}), it can easily be checked that $\dot\rho(t)\neq0$. This incorrectly implies that the system will evolve even if it is initially in the ground state. 

As noted earlier, the discrepancy in the SME arises because it assumes that the qubit-oscillator coupling plays no role in its dissipator. The dissipator of the SME predicts the zero temperature steady state of the system to be $\ket{g0}$, where $\ket{g}$ is the ground state of the qubit, ($\sigma_z\ket{e}=\ket{e}$ and $\sigma_z\ket{g}=-\ket{g}$). We can quantify the difference between the predictions of the SME and DME by calculating the distance, $d$, between the states $\ket{\Psi_0^-}$ and $\ket{g0}$: 
\begin{align}
d&=1-|\scpr{\Psi_0^-}{g0}|,\nonumber\\
&=\beta^2/2.
\end{align}
We see from the above equation that as the coupling strength becomes small, the distance $d$ tends to zero and thus the prediction of the SME starts to coincide with the prediction of the DME. This agrees with the common notion \cite{Carmichael-73, Cresser-92, Zoubi-00,Scala-07,Beaudoin-11} that the applicability of the SME is restricted only to the parameter domain where the coupling between the qubit and the oscillator is small.  

Similar to the derivation of the master equation in the zero temperature case, the master equation for the case when the reservoirs are at finite temperatures can also derived. For simplicity, we do not explicitly discuss the finite temperature case here and only give the results. If the qubit and the oscillator baths are at temperature $T$, the steady state of the DME can be calculated to be
\begin{align}
\rho_{ss}=\frac{e^{-H_{AD}/k_BT}}{\mathrm{Tr}\left[e^{-H_{AD}/k_BT}\right]},
\end{align}
where $k_B$ is the Boltzmann constant. This steady state correctly predicts the occupation probabilities of the dressed states of the coupled system to be the canonical Boltzmann distribution. It can also be shown that the finite temperature DME satisfies the principle of detailed balance whereas the SME does not.


\section{Qubit energy level fluctuations}\label{s.Dephasing}


The energy level splitting of a two level system can be sensitive to the environment with which it interacts. The effect of energy level fluctuations can be accounted for by introducing the following fluctuating Hamiltonian:
\begin{align}
H_{f}(t)=\hbar f(t)\sigma_z,
\end{align}  
where $f(t)$ is a stochastic term. Thus, in the presence of energy level fluctuations of the qubit, the total Hamiltonian governing the dynamics of the qubit-oscillator system becomes: 
\begin{align}
H(t)=H_{AD}+H_{f}(t).
\end{align} 
With the above Hamiltonian, it is now a question as to what is the ensemble averaged effect of $H_{f}(t)$ on the evolution of the coupled qubit-oscillator system. In Appendix \ref{a.Dephasing}, we find that in the presence of the fluctuating Hamiltonian, the evolution of the system is governed by:
\begin{align}\label{e.evol}
\dot{\rho}(t)=\frac{1}{i\hbar}\left[H_{AD},\rho(t)\right]-\frac{\gamma_f}{2}\left[S_z,\left[S_z,\rho(t)\right]\right],
\end{align}
where we have converted back to the Schr\"odinger picture from the interaction picture that was used in Eq. (\ref{a.evol_deph}). The operator, $S_z$, is the dephasing operator which has the form:
\begin{align}
S_z=\sum_{N}\scpr{N_+}{N_-}\left(\ket{\Psi_N^+}\bra{\Psi_N^+}-\ket{\Psi_N^-}\bra{\Psi_N^-}\right).
\end{align}

To understand the dynamical implications of Eq. (\ref{e.evol}), we look at the evolution of the various matrix elements of the coupled qubit-oscillator density matrix:
\begin{align}
\matel{\Psi_N^{\pm}}{\dot\rho(t)}{\Psi_M^{\pm}}=&-\frac{\gamma_f}{2}\left(\scpr{N_+}{N_-}-\scpr{M_+}{M_-}\right)^2\nonumber\\
&\times\matel{\Psi_N^{\pm}}{\rho(t)}{\Psi_M^{\pm}},\nonumber\\
=&-\frac{\gamma_f}{2}\mathcal{O}(\beta^4)\matel{\Psi_N^{\pm}}{\rho(t)}{\Psi_M^{\pm}},\nonumber\\
\approx&0,\label{e.deph_NM}\\
\matel{\Psi_N^{\mp}}{\dot\rho(t)}{\Psi_M^{\pm}}=&-\frac{\gamma_f}{2}\left(\scpr{N_+}{N_-}+\scpr{M_+}{M_-}\right)^2\nonumber\\
&\times\matel{\Psi_N^{\mp}}{\rho(t)}{\Psi_M^{\pm}}\label{e.deph_pm}.
\end{align}
We see from Eq. (\ref{e.deph_NM}) that the population of the various energy eigenstates of the qubit-oscillator system does not change because of the energy-level fluctuations of the qubit. This is in contrast to what is found in the analysis of qubit-oscillator evolution in the near resonant ultra-strong coupling regime where energy-level fluctuations of the qubit tend to induce population transfer between the eigenstates of the Rabi Hamiltonian \cite{Beaudoin-11}. We also notice from Eq. (\ref{e.deph_NM}) that qubit level fluctuations does not have any dynamical consequence if the population is entirely in the harmonic oscillator $\mathcal{HO}^+$ or entirely in $\mathcal{HO}^-$. From Eq. (\ref{e.deph_pm}) we see that any initial coherence between the various eigenstates of $\mathcal{HO}^+$ and $\mathcal{HO}^-$, i.e. between states $\ket{\Psi_{N}^+}$ and $\ket{\Psi_{M}^-}$, gets exponentially damped to zero. 


\section{Driven damped qubit-oscillator system}\label{s.Drive}


We now consider the case when the oscillator is pumped by a classical drive of frequency $\omega_p$ and amplitude $\Omega_p$. The Hamiltonian modeling the pump is:
\begin{align}\label{e.H_p}
H_{p}&=\hbar\Omega_p(a +a^{\dagger})(e^{i\omega_p t}+e^{-i\omega_p t}),\nonumber\\
&=\hbar\Omega_p(S +S^{\dagger})(e^{i\omega_p t}+e^{-i\omega_p t}),
\end{align}
where $S$ is the dressed lowering operator defined as
\begin{align}\label{e.S}
S&=\sum_{k>j}\ket{j}\bra{k}\matel{j}{a+a^\dagger}{k},\nonumber\\
&=a_-+a_+-2\beta\sum_{N}\ket{\Psi^-_N}\bra{\Psi^+_N}.
\end{align} 
In the above equation, we have used the matrix elements of $(a+a^\dagger)$ derived in Eq. (\ref{a.e.C_jk_osc}). 

Including the pump Hamiltonian and the effect of qubit energy level fluctuations in Eq. (\ref{e.master_zero}), we get the master equation describing the evolution of the pumped qubit-oscillator system to be:
\begin{align}\label{e.master_driven}
\dot{\rho}(t)=&\frac{1}{i\hbar}\left[ H_{AD},\rho(t) \right]+\frac{\hbar\Omega_p}{i\hbar}\left[ S +S^{\dagger},\rho(t) \right](e^{i\omega_p t}+e^{-i\omega_p t})\nonumber\\
&+\mathcal{J}_{osc}\rho(t)+\mathcal{J}_{qbit}\rho(t)-\frac{\gamma_f}{2}\left[S_z,\left[S_z,\rho(t)\right]\right],
\end{align}
In order to remove the explicit time dependence from the right hand side of Eq. (\ref{e.master_driven}), we go to a frame rotating at the pump frequency and define the density matrix in the rotating frame by $\chi(t)$:
\begin{align}
\chi(t)=e^{i\omega_p(a_+^{\dagger}a_++a_-^{\dagger}a_-)t}\rho(t)e^{-i\omega_p(a_+^{\dagger}a_++a_-^{\dagger}a_-)t}.
\end{align}
In this rotating frame, the term $\sum\ket{\Psi^-_N}\bra{\Psi^+_N}$ in the driving part of the Hamiltonian rotates at the pump frequency, $\omega_p$. In the spirit of the rotating wave approximation, we average this term to zero. Two other terms: $Se^{-i\omega_p t}$ and $S^{\dagger}e^{i\omega_p t}$, also average to zero. Thus the master equation in the rotating frame becomes:
\begin{align}\label{e.driven_mater_rwa}
\dot{\chi}(t)=&\mathcal{L}^{+}\chi(t)+\mathcal{L}^{-}\chi(t)\nonumber\\
&+\kappa\sum_{N}\mathcal{D}\left[\ket{\Psi_{N}^{-}}\bra{\Psi_{N}^{+}}\right]\chi(t)\nonumber\\
&-\frac{\gamma_f}{2}\left[S_z,\left[S_z,\chi(t)\right]\right],
\end{align}
where 
\begin{align}
\mathcal{L}^{\pm}\chi(t)=&-i\Delta_{\pm}\left[a_\pm^{\dagger}a_\pm,\chi(t)\right]-i\Omega_p\left[ a_\pm+a_\pm^{\dagger},\chi(t)\right]\nonumber\\
&+\Gamma\mathcal{D}\left[a_\pm\right]\chi(t),\nonumber\\
\kappa=&\gamma(\tilde{\omega}_N)+4\beta^2\Gamma(\tilde{\omega}_N),\nonumber\\
\Delta_\pm=&\omega_\pm-\omega_p.
\end{align}
In the above equations, we have assumed that $\Gamma(\omega_+)=\Gamma(\omega_-)=\Gamma$ and $\gamma(\tilde{\omega}_N)+4\beta^2\Gamma(\tilde{\omega}_N)=\kappa$ does not depend on $N$ for simplicity. The Liouvillians $\mathcal{L}^{+}$ and $\mathcal{L}^{-}$ generate evolution of the harmonic oscillators $\mathcal{HO}^{+}$ and $\mathcal{HO}^{-}$ respectively. The second-last term in Eq. (\ref{e.driven_mater_rwa}) transfers population incoherently from $\mathcal{HO}^{+}$ to $\mathcal{HO}^{-}$ at the rate $\kappa$. 

The structures of the Liovillians, $\mathcal{L}^+$ and $\mathcal{L}^-$, are formally the same as that of a Liouvillian that generates evolution of a bare driven damped harmonic oscillator \cite{Breur-Petruccione}. The steady states generated by $\mathcal{L}^\pm$ are coherent states, $\ket{\alpha_\pm}$, and are easily evaluated to be:
\begin{align}\label{e.alpha_m}
\ket{\alpha_{\pm}}&=e^{-\frac{|\alpha_\pm|^2}{2}}\sum_N\frac{(\alpha_\pm)^{N}}{\sqrt{N!}}\ket{\Psi_N^\pm},\nonumber\\
\alpha_\pm&=\frac{\Omega_p}{i\frac{\Gamma}{2}-\Delta_\pm}.
\end{align}
The fact that $\ket{\alpha_\pm}$ are indeed the steady states of $\mathcal{L}^{\pm}$ can be checked by verifying that $\mathcal{L}^{\pm}\ket{\alpha_{\pm}}=0$. We reiterate the important point that these steady states are coherent states in the dressed basis, $\{\ket{\Psi_{N}^\pm}\}$.

The long term behavior of the system dynamics can be understood by looking at the evolution of the following terms:
\begin{align}
\sum_{N}\matel{\Psi_N^+}{\dot{\chi}(t)}{\Psi_N^+}&=-\kappa\sum_{N}\matel{\Psi_N^+}{\chi(t)}{\Psi_N^+},\label{e.pchip}\\
\sum_{N}\matel{\Psi_N^-}{\dot{\chi}(t)}{\Psi_N^-}&=\kappa\sum_{N}\matel{\Psi_N^+}{\chi(t)}{\Psi_N^+}.\label{e.mchim}
\end{align}
In the above expressions, we have used Eq. (\ref{e.deph_NM}). The left hand side of Eq. (\ref{e.pchip}) corresponds to the total population of the system within the $\mathcal{HO}^+$ oscillator. Similarly, the left hand side of Eq. (\ref{e.mchim}) corresponds to the total population within the $\mathcal{HO}^-$ oscillator subspace.
From Eqs. (\ref{e.pchip}) we see that in the long time limit, 
\begin{align}
\lim_{t\kappa\gg1}\sum_{N}\matel{\Psi_N^+}{\dot{\chi}(t)}{\Psi_N^+}=0.
\end{align}
Thus, in the steady state, there is no population in the oscillator, $\mathcal{HO}^+$. From Eq. (\ref{e.mchim}), we see that all the initial population in $\mathcal{HO}^+$ is incoherently transferred to $\mathcal{HO}^-$. We conclude from these considerations that in the limit, $\kappa t\gg 1$, all the population is found only within the $\mathcal{HO}^-$ oscillator subspace. As noted in Eq. (\ref{e.alpha_m}), the steady state within the $\mathcal{HO}^-$ oscillator subspace is $\ket{\alpha_-}$. Thus in the long time limit, any initial state will evolve to the steady state
\begin{align}\label{e.chiss}
\chi_{ss}=\ket{\alpha_-}\bra{\alpha_-}.
\end{align} 
Defining the sum of average excitation numbers of the $\mathcal{HO}^-$ and $\mathcal{HO}^+$ oscillators to be 
\begin{align}
\mathcal N(t)&=\la a_-^\dagger(t) a_-(t)\ra+\la a_+^\dagger(t) a_-(t)\ra,\nonumber\\
&=\mathcal N^-(t)+\mathcal N^+(t),
\end{align}
we see from Eq. (\ref{e.chiss}) that in the steady state, we have 
\begin{align}\label{e.N_ssm}
 \mathcal{N}_{ss}&=\mathcal N_{ss}^-,\nonumber\\
&=|\alpha_-|^2. 
\end{align}


\section{Spectroscopy}\label{s.Spectroscopy}


In this section, we will see how one can use the sum of average excitation numbers, $\mathcal N_{ss}$, to experimentally find out the energy eigenstates of the composite qubit-oscillator system.  

The bath modes coupled to the system allow the system energy eigenstates to be probed through spectroscopy. In the physical scenario where the oscillator bath mode is used as a measurement channel, the output signal will depend upon the oscillator bath operator: $b_\lambda(t)$. From the system-bath interaction Hamiltonian given in Eq. (\ref{a.e.BS}), it is clear that the oscillator bath operator evolves as:
\begin{align}
b_\lambda(t)e^{i\nu_\lambda t}=&b_\lambda(0)-i h_\lambda\int_{0}^{t}\mathrm{dt'}\, S(t')e^{i\nu_\lambda t'}.
\end{align}    
where $S(t)$ is defined in Eq. (\ref{e.S}).

\begin{figure}[t!]
\centering
\includegraphics[width=8cm, height=8cm]{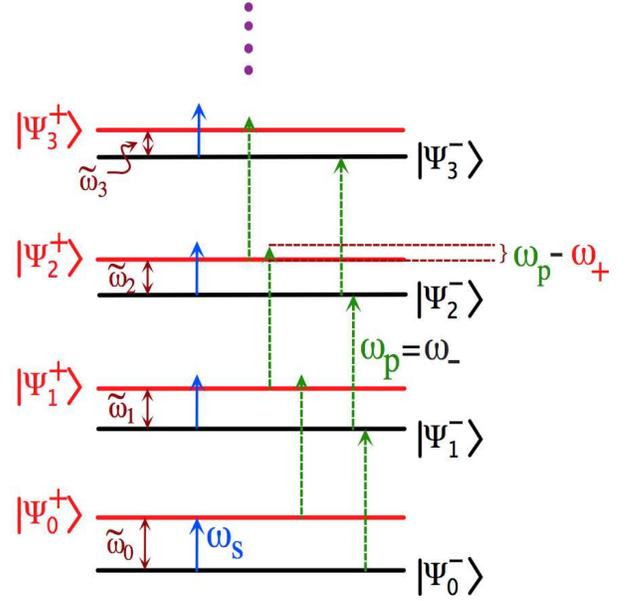}
\caption{Two-tone spectroscopy to probe the $\ket{\Psi_N^-}\leftrightarrow\ket{\Psi_N^+}$ transitions. The dashed (green) arrows indicate the pump frequency which is taken to be resonant with $\mathcal{HO}^-$, $\omega_p=\omega_-$. This pump is off-resonant with the $\mathcal{HO}^+$ oscillator, $\omega_p\neq\omega_+$. The solid (blue) arrows indicate the spectroscopy drive which is scanned around $\omega_0$. Whenever $\omega_s=\tilde\omega_N$, population gets transferred from $\ket{\Psi_N^-}$ to $\ket{\Psi_N^+}$. }
\label{fig:SpecLevelDiag}
\end{figure}
The rate at which the system dissipates energy into the oscillator bath is an experimentally measurable quantity. If the bath is initially in the vacuum state, the power that is dissipated by the system into the oscillator's bath in the steady state is given by
 \begin{align}\label{e.P_ss}
P_{ss}&=\lim_{t\kappa,t\Gamma\gg1}\frac{1}{t}\sum_{\lambda}\hbar\nu_{\lambda}\la b_\lambda^\dagger(t)b_\lambda(t)\ra_{ss},\nonumber\\
&=\lim_{t\kappa,t\Gamma\gg1}\frac{\hbar}{t}\int_{0}^{\infty} d\nu g(\nu)h^2(\nu)\nu\nonumber\\
&\times\int_{0}^{t} dt_1 \int_{0}^{t} dt_2 \la S^{\dagger}(t_1)S(t_2)\ra_{ss}   e^{-i\nu(t_2-t_1)}.
\end{align}
The sum over modes, $\lambda$, has been converted into an integral with $g(\nu)$ being the density of states. Let us suppose that the oscillator is driven by a classical pump of frequency $\omega_p$ which is close to the bare oscillator frequency $\omega$. In this case, as in the previous section, the term $\sum\ket{\Psi^-_N}\bra{\Psi^+_N}$ in the operator $S(t)$ oscillates at $\omega_p$ and can be averaged to zero. Following the usual approach \cite{Cohen-77} of using the Markov approximation to calculate the integral in Eq. (\ref{e.P_ss}), we get:
\begin{align}
P_{ss}&=\hbar\omega_p\frac{\Gamma(\omega_p)}{2}\left(\la a_{+}^{\dagger}a_{+}\ra_{ss}+\la a_{-}^{\dagger}a_{-}\ra_{ss}\right),\nonumber\\
&=\hbar\omega_p\frac{\Gamma(\omega_p)}{2}\mathcal N_{ss}.
\end{align}
From the above equation, we see that the experimentally measurable quantity, $P_{ss}$, is proportional to the sum of the steady state average excitations of the $\mathcal{HO}^+$ and the $\mathcal{HO}^-$ oscillators. 

We now discuss how one can use $P_{ss}$ to probe the energy spacing between the dressed states $\ket{\Psi_N^+}$ and $\ket{\Psi_N^-}$, i.e. states having the same values of $N$. For this purpose, we propose a spectroscopic technique that is similar to the one discussed in \cite{Gambetta-06}. The technique requires driving the oscillator by a pump drive with frequency $\omega_p$ and a spectroscopy drive with frequency $\omega_s$. The pump drive is taken to be resonant with the oscillator $\mathcal{HO}^-$, $\omega_p=\omega_-$, and the frequency of the spectroscopy drive is scanned near the qubit frequency, $\omega_s\sim\omega_0$. The setup for this two-tone spectroscopy is shown in Fig. (\ref{fig:SpecLevelDiag}).

\begin{figure}[t!]
\includegraphics[width=9cm, height=4.5cm]{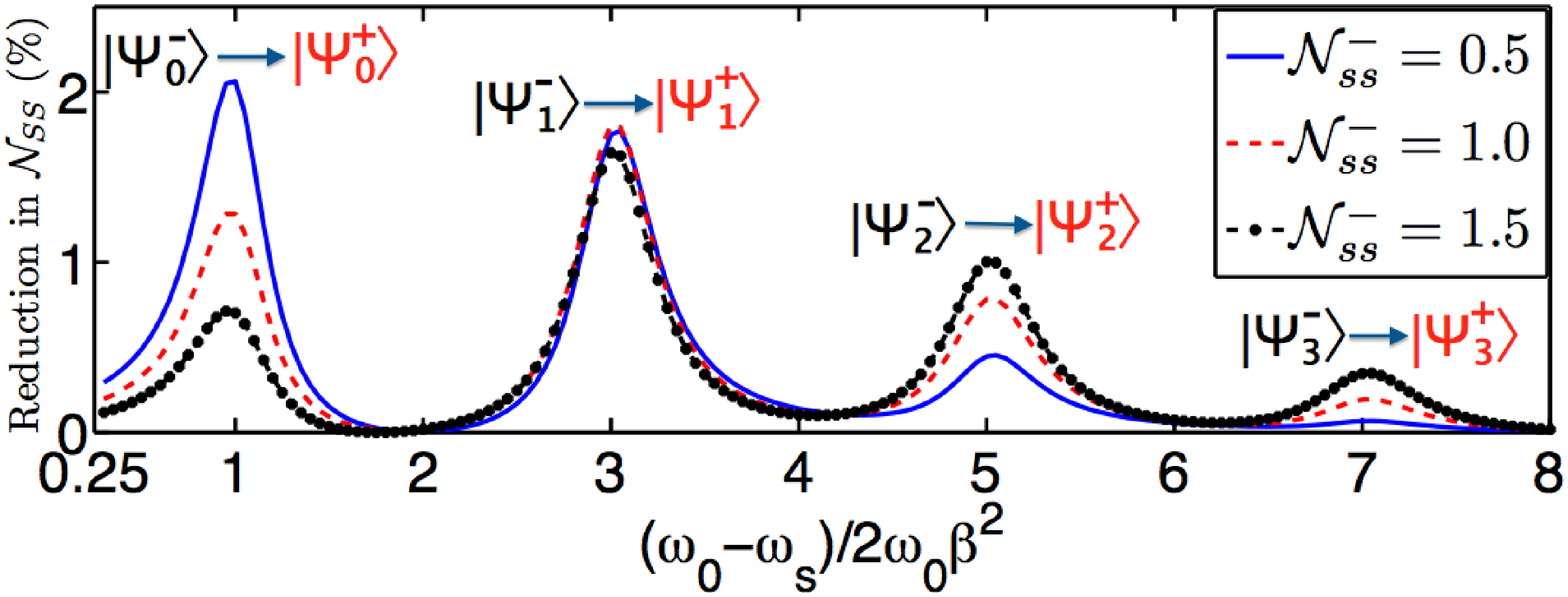}
\caption{Numerically evaluated percentage reduction in the steady state excitation number, $\mathcal N_{ss}$, as a function of the spectroscopy drive frequency, $\omega_s$. The pump frequency is taken to be resonant with the $\mathcal{HO}^-$ oscillator, $\omega_p=\omega_-$. The qubit frequency and the coupling parameter is taken to be $\omega_0=0.3\omega$ and $\beta=0.1$ respectively. The damping rates are taken to be $\Gamma=6\kappa=3\gamma_f=3\omega_0\beta^2/5$. The spectroscopy drive strength is taken to be $\Omega_s=\kappa$. Maximum reduction in the power is achieved whenever $\omega_0-\omega_s=2\omega_0\beta^2(2N+1)$. The various curves correspond to various pump strengths which are parameterized by the corresponding average excitation numbers, $\mathcal N_{ss}^-$, calculated in the absence of the spectroscopy drive. For higher average excitation numbers, peaks with higher values of $N$ in Eq. (\ref{e.NN_Res}) start getting more dominant (this is easily seen in the figure for $(\omega_0-\omega_s)/2\omega_0\beta^2=5$), whereas peaks at lower values of $N$ get smaller.}
\label{fig:Diff_Pump}
\end{figure}
In the absence of the spectroscopy drive, the pump drive forces the system to oscillate and in the steady state, when all the population relaxes to the $\mathcal{HO}^-$ oscillator, the average excitation number is given by Eq. (\ref{e.N_ssm}) with $\Delta_-=0$ (since $\omega_p=\omega_-$):
\begin{align}\label{e.Nssm}
\mathcal N_{ss}=\mathcal N_{ss}^-=4\Omega^2_p/\Gamma^2.
\end{align}

Now in the presence of the spectroscopy drive, whenever the $\ket{\Psi_N^-}\leftrightarrow\ket{\Psi_N^+}$ transition becomes resonant with the spectroscopy drive frequency, i.e. when
\begin{align}\label{e.NN_Res}
&\omega_s=\omega_0(1-2\beta^2-4N\beta^2),\nonumber\\
\implies&\frac{(\omega_0-\omega_s)}{2\omega_0\beta^2}=2N+1, 
\end{align}
appreciable population gets transferred from the $\mathcal{HO}^-$ oscillator to the $\mathcal{HO}^+$ oscillator. This causes a reduction in the excitation number of $\mathcal{HO}^-$. Since the pump drive is off-resonant with the $\mathcal{HO}^+$ oscillator, $\omega_p\neq\omega_+$, the population that gets transferred does not get resonantly driven. Hence, the reduction in the excitation number of the $\mathcal{HO}^-$ oscillator does not get compensated by the increase in excitation number of the $\mathcal{HO}^+$ oscillator. This results in a net reduction of $\mathcal N_{ss}$ whenever the spectroscopy drive frequency, $\omega_s$, satisfies the resonance condition given in Eq. (\ref{e.NN_Res}). Reduction in the average excitation number manifests itself through a reduction in the dissipated power. Thus, by monitoring the decrease in the dissipated power, $P_{ss}$, as the spectroscopy drive is scanned around $\omega_0$, one can determine the various frequencies, $\tilde\omega_N=\omega_0(1-2\beta^2-4N\beta^2)$, that are associated with the transitions between the dressed states $\ket{\Psi_N^-}$ and $\ket{\Psi_N^+}$. 

In order to include the effect of the spectroscopy drive, we add to the master equation (\ref{e.master_driven}) a spectroscopy drive Hamiltonian, $H_s$, which is identical to the pump Hamiltonian, $H_p$ given in Eq. (\ref{e.H_p}), but with $\Omega_p$ and $\omega_p$ replaced by spectroscopy drive strength, $\Omega_s$, and frequency, $\omega_s$, respectively. We then numerically solve the master equation and evaluate the steady state average excitation number $\mathcal N_{ss}$. We plot in Fig. \ref{fig:Diff_Pump} the percentage reduction in $\mathcal N_{ss}$ as a function of the spectroscopy drive frequency, $\omega_s$. In agreement with our understanding, we clearly see from the figure that whenever the resonance condition defined in Eq. (\ref{e.NN_Res}) is satisfied, there is a reduction in the the average excitation number. This provides a clear signature using which one can probe the transition energies between the states $\ket{\Psi_N^+}$ and $\ket{\Psi_N^-}$.

In Fig. \ref{fig:Diff_Pump}, we also study the effect of the strength of the pump drive on the reduction of the steady state power. Using Eq. (\ref{e.Nssm}), we see that the pump strength, $\Omega_p$, can be parameterized by $\mathcal N_{ss}^-$, which is the average excitation number of the $\mathcal{HO}^-$ oscillator in the absence of the spectroscopy drive. Since the pump is taken to be a coherent drive, one expects that in the steady state, the populations in the $\ket{\Psi_N^-}$ states will have a Poisson distribution. As $\mathcal N_{ss}^-$ increases, population in the higher $\ket{\Psi_N^-}$ states also increases. This increases the ability of the spectroscopy drive to transfer population to the $\ket{\Psi_N^+}$ states with higher values of $N$. Thus, we expect the peaks with higher values of $N$ to be more dominant as $N_{ss}^-$ increases. This is in agreement with the plots in Fig. \ref{fig:Diff_Pump} where we see that as $\mathcal N_{ss}^-$ increases, the peaks with higher values of $N$ in the transition $\ket{\Psi_N^-}\rightarrow\ket{\Psi_N^+}$ start getting more pronounced while peaks with lower values of $N$ get smaller. 



\section{Region of Validity}\label{s.Validity}


The eigenvalues, eigenfunctions and the Lindblad master equation derived in this report are based on certain approximations. For a better understanding of the validity of these analytic formulas, we list below the various constraints under which these results were derived:
\begin{itemize}
\item[](a) $\omega_0\ll\omega$: Quantifies the quasi-degeneracy of the qubit. Under this condition, the time scales associated with the qubit evolution and the oscillator evolution are well separated and one can conclude that the fast oscillator adiabatically follows the slow qubit and the fast oscillations of the qubit at or above the oscillator frequency can be averaged to zero. This validates the use of the adiabatic approximation (Sect. \ref{s.Ad-Approx}). The maximum possible qubit frequency within which the adiabatic approximation is valid is $\omega_0=0.3\omega$ \cite{Shantanu-12}. 
\item[](b) $\mathcal{O}(\beta^4)\sim 0$: Justifies the approximation made in Eq. (\ref{e.exp_approx}). Even within this approximation, the coupling can be strong enough, $|\beta_{max}|=0.2$, to invalidate the RWA. 
\item[](c) $N\ll(1/2\beta)^2$: Allows us to restrict the power series expansion of $L_N$ to the first two terms only. It is used in Eq. (\ref{e.lag_approx}) and sets a limit on the validity of relations (\ref{e.apm}).  
\item[](d) Secular approximation: Assumes that the energy difference between dressed states is much bigger than the relaxation rates between them (see Eq. (\ref{a.e.secular})). We have also used this approximation to remove the ``counter rotating terms", $B(t)S(t)$ and $B^\dagger(t)S^\dagger(t)$,  from the system-bath interaction Hamiltonian to arrive at Eq. (\ref{a.e.BS}). 
\item[](e) Born approximation: Assumes that the coupling of the qubit and the oscillator to their respective baths, $q_\lambda$ and $h_\lambda$, are weak. This allows restricting the system-bath interaction to second order in the coupling strength. Also within this approximation, one assumes that the reservoir state is not effected due to its interaction with the system. Born approximation is used in Eqs. (\ref{a.e.Born}).
\item[](f) Markov approximation: Assumes that the bath correlation time is much smaller than the time scale in which the system evolves. Markov approximation is used in Eqs. (\ref{a.e.Markov1}) and (\ref{a.e.fjklm}).
\end{itemize}


\section{Conclusion}\label{s.Conclusion}


In this report, we investigated the effect of the environment on the dynamics of a coupled qubit-oscillator system. We restricted our analysis to the parameter regime where the qubit is quasi-degenerate, $\omega_0\ll\omega$, and the coupling is allowed to be ultra-strong. For such a parameter regime, the rotating wave approximation breaks down. Also in this parameter domain of interest, the standard master equation which is commonly used to describe the effect of the environment on the qubit-oscillator system is known to fail. For example, when the baths are taken to be at zero temperature, the SME incorrectly predicts the ground state of the composite qubit-oscillator system to be not the equilibrium state. 

To get around the nonphysical predictions of the SME, we microscopically derived the dressed state master equation. The DME is shown to predict the correct thermal equilibrium state. We showed that the differences between the predictions of the SME and the DME decreases as the coupling strength decreases. This is in agreement with the fact that the SME is valid only when the coupling is small but must be abandoned when the coupling is ultra-strong. 

For deriving the DME, we needed to find the eigen-energies and eigen-functions of the Rabi Hamiltonian. We achieved this by using the \textit{adiabatic approximation}. The adiabatic approximation is valid only within the quasi-degenerate qubit regime. It was shown that the eigenstructure of the qubit-oscillator system comprises of two sets of harmonic oscillators, $\mathcal{HO^\pm}$. These oscillators have different frequencies, $\omega_\pm=\omega\mp2\omega_0\beta^2$, and different ground state energies. 

The damping rates, $\Gamma(\omega_{\pm})$, $\Gamma(\tilde\omega_N)$ and $\gamma(\tilde\omega_N)$, entering the DME depend upon the bath density of states evaluated at the eigen-frequencies of the coupled qubit-oscillator system. Since the eigen-frequencies depend upon the coupling strength, $\beta$, it is in principle possible to change the damping rates just by tuning the coupling. This dependence of damping rates on the coupling strength does not arise within the SME. 

The master equation governing the dynamics of the damped qubit-oscillator system when the oscillator is driven coherently is derived in Sec. \ref{s.Drive}. When the drive frequency is taken to be nearly the same as the bare oscillator frequency, the steady state of the driven damped system is found to be a coherent state in the subspace spanned by the oscillator, $\mathcal{HO^-}$. 

A two tone spectroscopic technique which can probe the energy differences between the eigenstates $\ket{\Psi_N^+}$ and $\ket{\Psi_N^-}$ is discussed in Sec. \ref{s.Spectroscopy}. By solving the dressed master equation, we find that whenever the spectroscopic drive frequency becomes resonant with the $\ket{\Psi_N^-}\longrightarrow\ket{\Psi_N^+}$ transition, the steady state power dissipated by the system, $P_{ss}$, reduces. This provides a tool with which one can probe the eigenenergies of the composite qubit-oscillator system in the quasi-degenerate qubit and ultra-strong coupling regime where the RWA breaks down.

\begin{acknowledgments}
We thank A. Vigoren, C.J. Broadbent and F. Beaudoin for important remarks and helpful discussions. Financial support was received from NSF PHY-1203931. 
\end{acknowledgments}


\appendix


\section{Master equation}\label{a.Master_equation}

In this appendix, we will derive the master equation for the qubit-oscillator system when the qubit and the oscillator are interacting with separate bosonic baths. The analysis will be based on Born-Markov and secular approximations and will be carried out in the dressed basis. We will assume that the initial state of each bath is a thermal state. We explicitly carry the calculations only for the case when the bath is at zero temperature. 

We first consider the case in which only the oscillator is coupled to its corresponding bath. For this setup, the system-bath interaction Hamiltonian is given by: 
\begin{align}
H^{I}_{osc}&=\hbar\sum_{\lambda}h_{\lambda} (b^{\dagger}_{\lambda}+b_{\lambda})C,
\end{align} 
where $C$ is a system operator that, according to Eq. (\ref{e.H_int}), is $(a^{\dagger}+a)$. We now write the interaction Hamiltonian in the interaction picture:
\begin{align}
\tilde{H}^{I}_{osc}(t)=\hbar\sum_{\lambda,(j,k)}h_{\lambda} (b^{\dagger}_{\lambda}e^{i\nu_{\lambda}t}+b_{\lambda}e^{-i\nu_{\lambda}t})C_{jk}\ket{j}\bra{k}e^{-i\Delta_{kj}t},
\end{align}
where the set of states $\{\ket{j}\}$ are the eigenstates of the coupled qubit-oscillator system as given in Eq. (\ref{e.adiabatic_eigen}). The coefficients, $C_{jk}$, and the frequencies, $\Delta_{kj}$, are defined as:
\begin{align}
\Delta_{kj}=\omega_{k}-\omega_{j}\nonumber\\
C_{jk}=\matel{j}{C}{k}.
\end{align}

In order to aid making the secular approximation, we write the interaction Hamiltonian in a more convenient form:
\begin{align}
\tilde{H}^{I}_{osc}(t)&=\hbar\sum_{\lambda,(k>j)}h_{\lambda} (b^{\dagger}_{\lambda}e^{i\nu_{\lambda}t}+b_{\lambda}e^{-i\nu_{\lambda}t})C_{jk}\ket{j}\bra{k}e^{-i\Delta_{kj}t}\nonumber\\
&+\hbar\sum_{\lambda,(k<j)}h_{\lambda} (b^{\dagger}_{\lambda}e^{i\nu_{\lambda}t}+b_{\lambda}e^{-i\nu_{\lambda}t})C_{jk}\ket{j}\bra{k}e^{-i\Delta_{kj}t}\nonumber\\
&+\hbar\sum_{\lambda,(j)}h_{\lambda} (b^{\dagger}_{\lambda}e^{i\nu_{\lambda}t}+b_{\lambda}e^{-i\nu_{\lambda}t})C_{jj}\ket{j}\bra{j}.
\end{align}
Taking $C=(a+a^{\dagger})$, it can be easily shown that $C_{jj}=0$. So, we get:
\begin{align}
\tilde{H}^{I}_{osc}(t)&=\hbar\sum_{\lambda,(k>j)}h_{\lambda} (b^{\dagger}_{\lambda}e^{i\nu_{\lambda}t}+b_{\lambda}e^{-i\nu_{\lambda}t})C_{jk}\ket{j}\bra{k}e^{-i\Delta_{kj}t}\nonumber\\
&+H.C,\nonumber\\
&=(B(t)+B^{\dagger}(t))(S(t)+S^{\dagger}(t)),
\end{align}
where we have defined 
\begin{align}\label{a.e.Bath_System}
B(t)&=\hbar\sum_{\lambda}h_{\lambda} b_{\lambda}e^{-i\nu_{\lambda}t},\nonumber\\
S(t)&=\sum_{(k>j)}C_{jk}\ket{j}\bra{k}e^{-i\Delta_{kj}t}.
\end{align}
It should be noted that $B(t)$ is the lowering operator for the bath and $S(t)$ is the analogue of the system lowering operator in the dressed basis. According to the first secular approximation, we neglect the terms $B(t)S(t)$ and $B(t)^{\dagger}S(t)^{\dagger}$ from the interaction Hamiltonian to get:
\begin{align}\label{a.e.BS}
\tilde{H}^{I}_{osc}(t)&=B(t)S^{\dagger}(t)+B^{\dagger}(t)S(t).
\end{align}

Within the usual Born approximation of treating the system-bath interaction only up to second order in the coupling strength, $h_l$, and assuming the reservoir state to be unchanged during the dynamics, the evolution of the reduced density matrix of the qubit-oscillator system in the interaction picture follows the following integro-differential equation:
\begin{align}\label{a.e.Born}
\dot{\tilde{\rho}}(t)&=\frac{\Tr{B}}{(i\hbar)^2}\int_{0}^{t} dt'\left[\tilde{H}_{I}(t)\left[\tilde{H}_{I}(t'),\tilde{\rho}(t')\otimes\rho_{B}(0)\right]\right].
\end{align}
In Eq. (\ref{a.e.Born}), $\tilde{\rho}(t)$ is the reduced density matrix of the qubit-oscillator system in the coupled system interaction picture, $\Tr{B}$ refers to taking the trace over the bath degrees of freedom and $\rho_B(0)$ is the initial density matrix of the bath. For the zero temperature result, $\rho_B(0)$ is taken to be the vacuum state. Putting the interaction Hamiltonian, Eq. (\ref{a.e.BS}), in Eq. (\ref{a.e.Born}) and collecting all the non zero terms that contribute to the evolution, we get:
\begin{align}\label{a.e.master_interaction}
\dot{\tilde{\rho}}(t)&=\frac{1}{(i\hbar)^2}\int_{0}^{t} dt' \Big( S^{\dagger}(t)S(t')\tilde{\rho}(t')\la B(t)B^{\dagger}(t')\ra\nonumber\\
&-S(t)\tilde{\rho}(t')S^{\dagger}(t')\la B(t')B^{\dagger}(t)\ra+h.c.\Big),
\end{align} 
where $\la B(t)B^{\dagger}(t')\ra\equiv\Tr{B}\left\{B(t)B^{\dagger}(t')\rho_{B}(0)\right\}$ is the bath correlation function. From Eq. (\ref{a.e.Bath_System}) it can easily be seen that $\la B(t)B^{\dagger}(t')\ra=\la B(t-t')B^{\dagger}(0)\ra$. Let us now evaluate the first integral:
\begin{align}\label{a.e.I1}
I_{1}(t)&=\frac{1}{\hbar^2}\int_{0}^{t} dt'  S^{\dagger}(t)S(t')\tilde{\rho}(t')\la B(t)B^{\dagger}(t')\ra,\nonumber\\
&=\frac{1}{\hbar^2}\int_{0}^{t} d\tau  S^{\dagger}(t)S(t-\tau)\tilde{\rho}(t-\tau)\la B(\tau)B^{\dagger}(0)\ra.
\end{align}

Based on the assumption that the bath correlation function, $\la B(\tau)B^{\dagger}(0)\ra$, is peaked only for an interval $\tau=\tau_B$ which is much smaller than the relaxation time of the system, $\tau_R$, one can make the Markov approximation and replace $\tilde{\rho}(t-\tau)$ in Eq. (\ref{a.e.I1}) by $\tilde{\rho}(t)$ to get: 
\begin{align}\label{a.e.Markov1}
I_{1}(t)=\frac{1}{\hbar^2}\int_{0}^{t} d\tau  S^{\dagger}(t)S(t-\tau)\tilde{\rho}(t)\la B(\tau)B^{\dagger}(0)\ra.
\end{align}
Using Eq. (\ref{a.e.Bath_System}), we get: 
\begin{align}
I_{1}(t)=\sum_{k>j,m>l}C_{jk}C_{lm}^*\ket{m}\scpr{l}{j}\bra{k}\tilde{\rho}(t)f_{jklm}(t),
\end{align}
where
\begin{align}
f_{jklm}(t)=e^{i(\Delta_{ml}-\Delta_{kj})t}\sum_{\lambda}h_{\lambda}^2\int_{0}^{t} d\tau e^{-i(\nu_{\lambda}-\Delta_{kj})\tau}.
\end{align}
Converting the sum over the various bath modes in the above expression to an integral, we get:
\begin{align}
f_{jklm}(t)&=e^{i(\Delta_{ml}-\Delta_{kj})t}\int_{0}^{\infty} d\nu g(\nu)h^2(\nu)\nonumber\\
&\qquad\times\int_{0}^{t} d\tau e^{-i(\nu-\Delta_{kj})\tau},
\end{align}
where $g(\nu)$ is the density of modes for the bath. Assuming the relaxation time of the system to be much longer than the correlation time of the bath, the upper limit of the $\tau$ integral can be set to infinity. Neglecting the divergent frequency shift, $f_{jkj'k'}(t)$ can then be evaluated to get:
\begin{align}\label{a.e.fjklm}
f_{jklm}(t)&=\frac{\Gamma(\Delta_{kj})}{2}e^{i(\Delta_{ml}-\Delta_{kj})t} ,
\end{align}
where we have defined 
\begin{align}\label{a.e.Gamma}
\Gamma(\Delta_{kj})=2\pi g(\Delta_{kj})h^2(\Delta_{kj}),
\end{align}
the usual Fermi Golden Rule result. 

The terms in Eq. (\ref{a.e.fjklm}) that are oscillatory are additional non-secular terms and in the second secular approximation, these terms are averaged to zero. Thus any term, $f_{jklm}(t)$, for which $\Delta_{ml}\neq\Delta_{kj}$ is neglected. The secular approximation implies:
\begin{align}
f_{jklm}=\frac{\Gamma(\Delta_{kj})}{2}\delta(\Delta_{ml}-\Delta_{kj}),
\end{align}
where $\delta(\Delta_{ml}-\Delta_{kj})$ stands for the Kronecker delta function. This approximation is justified only if the relaxation rate of the system is much smaller than the oscillatory frequency \cite{Breur-Petruccione,Cohen-Reynaud}, i.e.:
\begin{align}\label{a.e.secular}
 \Gamma(\Delta_{kj})\ll|\Delta_{ml}-\Delta_{kj}|.
\end{align}
Thus, under the Born-Markov and secular approximations, one arrives at the following form for $I_1(t)$:
\begin{align}\label{a.e.I1_Cjk}
I_{1}(t)=&\sum_{k>j,m>l}C_{jk}C_{lm}^*\ket{m}\scpr{l}{j}\bra{k}\tilde{\rho}(t)\nonumber\\
&\times\frac{\Gamma(\Delta_{kj})}{2}\delta(\Delta_{ml}-\Delta_{kj}).
\end{align} 

We now look at the explicit form of the matrix elements, $C_{jk}$. In the quasi-degenerate qubit regime, the eigenstates of the qubit-oscillator system are:
\begin{align}
\ket{\Psi_{N}^{\pm}}&=\frac{1}{\sqrt{2}}\Big(\ket{+,N_+}\pm\ket{-,N_-}\Big).
\end{align} 
For $C=a+a^{\dagger}$, we get:
\begin{align}\label{a.e.C_jk_osc}
C_{N_{+}M_{+}}&=\matel{\Psi_{N}^{+}}{a+a^{\dagger}}{\Psi_{M}^{+}},\nonumber\\
&=\sqrt{M}\delta_{N,M-1}+\sqrt{M+1}\delta_{N,M+1},\nonumber\\
C_{N_{-}M_{-}}&=C_{N_{+}M_{+}},\nonumber\\
C_{N_{+}M_{-}}&=-2\beta\delta_{N,M},\nonumber\\
C_{N_{-}M_{+}}&=C_{N_{+}M_{-}}.
\end{align}
Putting Eq. (\ref{a.e.C_jk_osc}) in Eq. (\ref{a.e.I1_Cjk}), we get:
\begin{align}
I_{1}(t)&=\frac{1}{2}\Big(\Gamma(\omega_+)a^{\dagger}_+a_{+} +\Gamma(\omega_-)a^{\dagger}_-a_-\nonumber\\
&+4\beta^2\sum_{N}\Gamma(\tilde{\omega}_N)\ket{\Psi_{N}^{+}}\bra{\Psi_{N}^{+}}\Big)\tilde{\rho}(t).
\end{align}
where $\tilde{\omega}_N=\omega_0(1-4N\beta^2-2\beta^2)$.

Following the same arguments used to evaluate the integral $I_1(t)$, one can now evaluate the rest of the integrals appearing in Eq. (\ref{a.e.master_interaction}). In particular we have:
\begin{align}
I_2(t)&=\frac{1}{\hbar^2}\int_{0}^{t} dt' S(t)\tilde{\rho}(t')S^{\dagger}(t')\la B(t')B^{\dagger}(t)\ra,\nonumber\\
&=\frac{1}{2}\Big(\Gamma(\omega_+)a_{+}\tilde{\rho}(t)a^{\dagger}_+ +\Gamma(\omega_-)a_-\tilde{\rho}(t)a^{\dagger}_-\nonumber\\
&+4\beta^2\sum_{N}\Gamma(\tilde{\omega}_N)\ket{\Psi_{N}^{-}}\bra{\Psi_{N}^{+}}\tilde{\rho}(t)\ket{\Psi_{N}^{+}}\bra{\Psi_{N}^{-}}\Big).
\end{align}  
Putting these integrals in Eq. (\ref{a.e.master_interaction}) and going from the interaction picture to the Schrodinger picture, we get:
\begin{align}
\dot{\rho}(t)&=\frac{1}{i\hbar}\left[ H,\rho(t) \right]+\Gamma(\omega_+)\mathcal{D}\left[a_+\right]\rho(t)\nonumber\\
&+\Gamma(\omega_-)\mathcal{D}\left[a_-\right]\rho(t)\nonumber\\
&+4\beta^2\sum_{N}\Gamma(\tilde{\omega}_N)\mathcal{D}\left[\ket{\Psi_{N}^{-}}\bra{\Psi_{N}^{+}}\right]\rho(t),\nonumber\\
&=\frac{1}{i\hbar}\left[ H,\rho(t) \right]+\mathcal{J}_{osc}\rho(t),
\end{align}
where the dissipator $\mathcal{D}$ is defined in Eq. (\ref{e.Dissipator}). The Lindbladian, $\mathcal{J}_{osc}$, describes what effect of the oscillator being coupled to a zero temperature bath.

In order to derive the evolution equation for the system when the qubit is also coupled to a bath, we note that the matrix elements of $\sigma_x$ in the energy eigenbasis are:
\begin{align}\label{a.e.Cjk_qubit}
C_{N_{+}M_{+}}&=\matel{\Psi_{N}^{+}}{\sigma_x}{\Psi_{M}^{+}},\nonumber\\
&=0,\nonumber\\
C_{N_{-}M_{-}}&=C_{N_{+}M_{+}},\nonumber\\
C_{N_{+}M_{-}}&=\delta_{N,M},\nonumber\\
C_{N_{-}M_{+}}&=C_{N_{+}M_{-}}.
\end{align}
Using Eq. (\ref{a.e.Cjk_qubit}) and under the Born-Markov and secular approximations, the master equation now takes the form:
\begin{align}\label{a.e.Master_qo}
\dot{\rho}(t)&=\frac{1}{i\hbar}\left[ H,\rho(t) \right]+\mathcal{J}_{osc}\rho(t)\nonumber\\
&+\sum_{N}\gamma(\tilde{\omega}_N)\mathcal{D}\left[\ket{\Psi_{N}^{-}}\bra{\Psi_{N}^{+}}\right]\rho(t),\nonumber\\
&=\frac{1}{i\hbar}\left[ H,\rho(t) \right]+\mathcal{J}_{osc}\rho(t)+\mathcal{J}_{qbit}\rho(t),
\end{align}     
In the above equation, the damping rate $\gamma(\omega)$ is defined as:
\begin{align}\label{a.e.gamma}
\gamma(\omega)=2\pi g'(\omega) q^2(\omega),
\end{align}
where $g'(\omega)$ is the density of states and $q(\omega)$ is the coupling strength of the qubit with the bath evaluated at the frequency $\omega$.
The master equation (\ref{a.e.Master_qo}) describes the system evolution for the case of the qubit and the oscillator coupled to zero temperature baths.


\section{Energy level fluctuation of qubit}\label{a.Dephasing}


To understand the effects of fluctuating energy-level splitting of the qubit, we look at the evolution of the qubit-oscillator density matrix generated by the Hamiltonian:
\begin{align}
H(t)=H_{AD}+H_f(t).
\end{align} 
Going to the interaction picture, the evolution equation has the form:
\begin{align}\label{a.liouville}
\dot{\tilde\rho}_f(t)=\frac{\hbar f(t)}{i\hbar}\left[\tilde{\sigma}_{z}(t),\tilde\rho_f(t)\right],
\end{align}
where $\tilde\rho_f(t)=e^{iH_{AD}t}\rho_f(t)e^{-iH_{AD}t}$ and $\tilde{\sigma}_{z}(t)=e^{iH_{AD}t}\sigma_ze^{-iH_{AD}t}$ are operators in the interaction picture. The subscript $f$ in $\tilde\rho_f(t)$ indicates the evolution of the density matrix for a particular realization of the fluctuating term $f(t)$. 

Formally integrating Eq. (\ref{a.liouville}), we get the following integro-differential equation:
\begin{align}
\dot{\tilde\rho}_f(t)=&-if(t)\left[\tilde\sigma_z(t),\tilde\rho_f(t)\right]\nonumber\\
&-\int_0^t\mathrm{dt'}\,f(t)f(t')\left[\tilde\sigma_z(t),\left[\tilde\sigma_z(t'),\tilde\rho_f(t')\right]\right].
\end{align}
Assuming that the fluctuations are small, we only consider the effect of $f(t)$ up to its second order. Moreover, we assume that the fluctuations of $f(t)$ is not effected by the dynamis of the qubit-oscillator system. These are equivalent to the Born approximations. Taking the ensemble average of the various realizations of $f(t)$, we get:
\begin{align}
\dot{\tilde\rho}(t)=&\overline{\dot{\tilde\rho}_f(t)},\nonumber\\
=&-i\overline{f(t)}\left[\tilde\sigma_z(t),\overline{\tilde\rho_f(t)}\right]\nonumber\\
&-\int_0^t\mathrm{dt'}\,\overline{f(t)f(t')}\left[\tilde\sigma_z(t),\left[\tilde\sigma_z(t'),\overline{\tilde\rho_f(t')}\right]\right],
\end{align}  
where the overhead bars indicates ensemble averages. 

The statistics of $f(t)$ is in general governed by the physical system under consideration. For simplicity, we assume that $f(t)$ has zero mean. Moreover, we assume that it is a stationary random process having the following spectral decomposition of its two time correlation function:
\begin{align}
\la f(t)f(t')\ra&=\la f(t-t')f(0)\ra,\nonumber\\
&=\frac{1}{2\pi}\int_{-\infty}^{\infty}\mathrm{d\nu}\,\mathcal{G}(\nu)e^{-i\nu(t-t')}.
\end{align}  
With these considerations, the evolution equation becomes:
\begin{align}\label{a.tilde_rho}
\dot{\tilde\rho}(t)=&-\frac{1}{2\pi}\int_0^t\mathrm{dt'}\,\int_{-\infty}^{\infty}\mathrm{d\nu}\,\mathcal{G}(\nu)e^{-i\nu(t-t')}\nonumber\\
&\qquad\qquad\times\left[\tilde\sigma_z(t),\left[\tilde\sigma_z(t'),\tilde\rho(t')\right]\right]
\end{align}

We now find out the matrix elements of $\tilde\sigma_z(t)$ in the eigenbasis, $\ket{\Psi_N^\pm}$, of the adiabatic Hamiltonian, $H_{AD}$. We find that
\begin{align}
\matel{\Psi_N^s}{\tilde\sigma_z(t)}{\Psi_M^p}_{\substack{
            N\neq M\\
            s,p=\pm}}&=\orderof\left( \beta^2\right),\label{a.sz_MN}\\
\matel{\Psi_N^\pm}{\tilde\sigma_z(t)}{\Psi_N^\mp}&=\orderof\left( \beta^2\right),\label{a.sz_pm}\\
\matel{\Psi_N^\pm}{\tilde\sigma_z(t)}{\Psi_N^\pm}&=\pm\scpr{N_+}{N_-}.
\end{align} 
In the evolution Eq. (\ref{a.tilde_rho}), the operator $\tilde\sigma_z(t)$ appears quadratically. This means that the contribution of terms in Eqs. (\ref{a.sz_MN}) and (\ref{a.sz_pm}) will be of fourth order in the coupling: $\orderof\left( \beta^4\right)$, and hence we set them to zero. With this in mind, we can approximate $\tilde\sigma_z(t)$ in Eq. (\ref{a.tilde_rho}) to be $S_z$, where:
\begin{align}
S_z=\sum_{N}\scpr{N_+}{N_-}\left(\ket{\Psi_N^+}\bra{\Psi_N^+}-\ket{\Psi_N^-}\bra{\Psi_N^-}\right).
\end{align}
It should be noted that $S_z$ is time independent. Under the Markov approximation, the evolution Eq. (\ref{a.tilde_rho}) can now be evaluated to be:
\begin{align}\label{a.evol_deph}
\dot{\tilde\rho}(t)=-\frac{\gamma_f}{2}\left[S_z,\left[S_z,\tilde\rho(t)\right]\right],
\end{align}
where $\gamma_f$ is the dephasing rate quantified by the zero frequency component of the two time correlation's spectral distribution: 
\begin{align}
\gamma_f=\mathcal{G}(0).
\end{align}


\section{Adiabatic approximation and dispersive regime  beyond RWA}\label{a.Comparison}


As mentioned in Section \ref{s.Validity}, our analysis is restricted to the quasi-degenerate qubit regime, $\omega_0<0.3\omega$, and where the coupling is no stronger than $|\beta_{max}|=0.2$. This implies that all our analysis is valid only in the range where the coupling is much smaller than the detuning: $\omega|\beta|\ll |\omega-\omega_0|$ and the detuning is comparable to the bare qubit and the oscillator frequencies: $\omega-\omega_0\sim |\omega+\omega_0|$. This parameter regime is called the dispersive regime beyond the RWA \cite{Zueco}. 

It was shown by Zueco, et al. \cite{Zueco}, that the eigenvalues of $H_{Rabi}$ in this parameter regime can be approximately found by using the Schrieffer-Wolff (SW) transformation. According to the SW transformation, one unitarily rotates $H_{Rabi}$ to an approximate diagonal Hamiltonian: $H_{SW}=e^{-S}H_{Rabi}e^{S}$, where $S$ is an anti-hermitian matrix (see Eq. (13) of \cite{Zueco}), to get:
\begin{align}
H_{SW}=&\hbar\omega_{0}\frac{\sigma_{z}}{2} + \hbar\omega a^{\dagger}a+\hbar\omega_0\Big(\frac{\omega^2\beta^2}{\omega_0^2-\omega^2}\Big)\sigma_z(a^{\dagger}+a)^2.
\end{align}
Note that unlike in the dispersive regime within the RWA \cite{Blais-etal}, the oscillator excitation number, $a^\dagger a$, is not a constant of motion. Corresponding to the two eigenvalues of $\sigma_z$, the eigen-energies of $H_{SW}$ can easily be calculated to be 
\begin{align}\label{e.energy_dis}
\tilde{E}_{N}^{\pm}=\pm\hbar\omega_{0}/2 +\hbar\tilde{\omega}_{\pm}(N+1/2),
\end{align}
where
\begin{align}\label{e.om_pm_dis}
\tilde{\omega}_{\pm}=\omega\Big(1\mp\frac{4\omega\omega_0\beta^2}{\omega^2-\omega_0^2}\Big)^{1/2}.
\end{align}

Unlike the adiabatic approximation, which works only if the qubit frequency is much smaller than the oscillator frequency, the validity of the SW transformation does not depend up on the sign of the detuning: $\omega_0-\omega$. Thus, Eqs. (\ref{e.energy_dis}) and (\ref{e.om_pm_dis}) are valid for a red detuned qubit, $\omega_0<\omega$, as well as for a blue detuned qubit, $\omega_0>\omega$. In the quasi-degenerate regime, we have $\omega_0\ll\omega$ and so we can approximate the frequencies $\tilde{\omega}_\pm$ in Eq. (\ref{e.om_pm_dis}) to:
\begin{align}
\tilde{\omega}_{\pm}&=\omega\mp2\omega_0\beta^2.
\end{align}
Using the above approximate expression for $\tilde{\omega}_{\pm}$ in Eq. (\ref{e.energy_dis}), and comparing it with Eq. (\ref{e.adiabatic_eigen}), we see that the eigen-energies calculated using the adiabatic approximation matches exactly with the eigen-energies derived using the SW transformation: $E_N^\pm=\tilde{E}_{N}^{\pm}$. 

\begin{figure}[t]
\centering
\subfigure{\includegraphics[width=4.25cm, height=2.8cm]{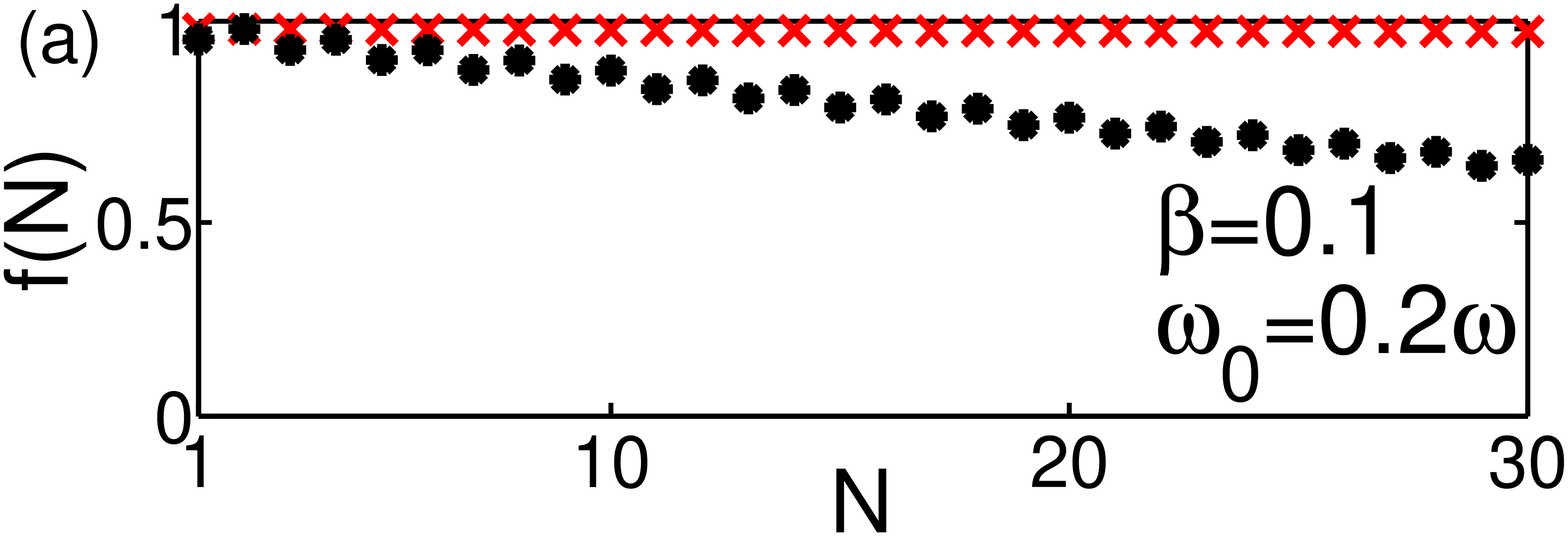}\label{fig:fidelity1}}
\subfigure{\includegraphics[width=4.25cm, height=2.8cm]{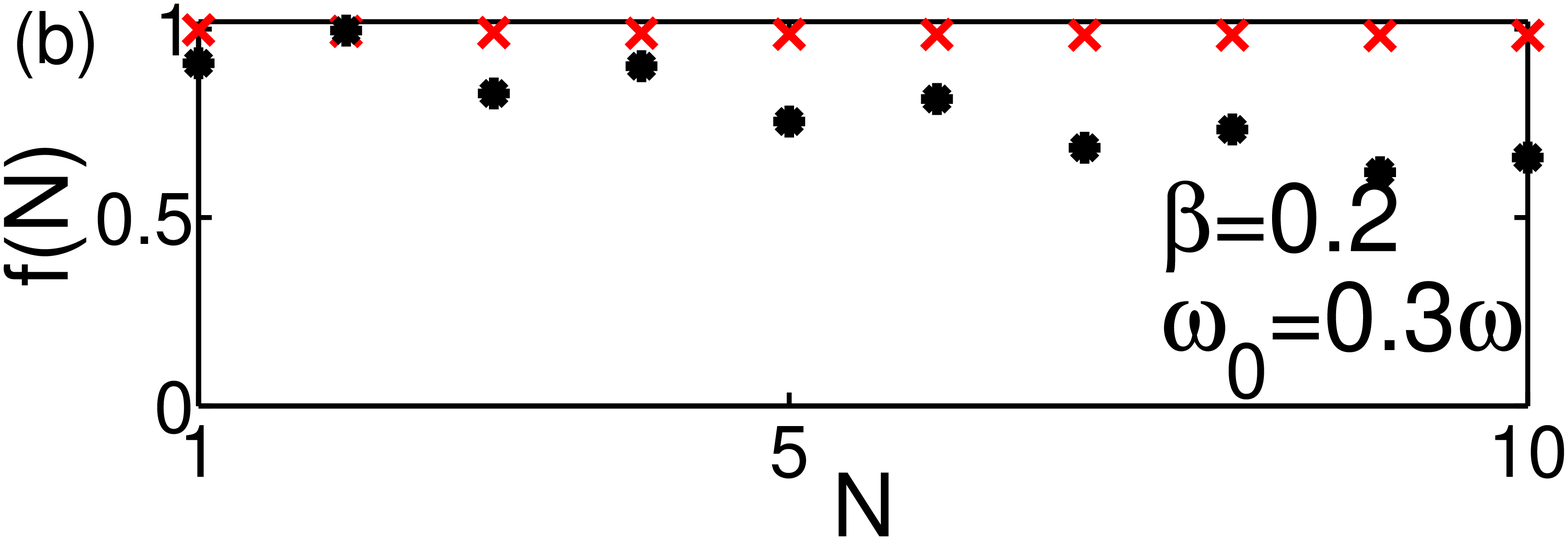}\label{fig:fidelity2}}
\subfigure{\includegraphics[width=4.25cm, height=2.8cm]{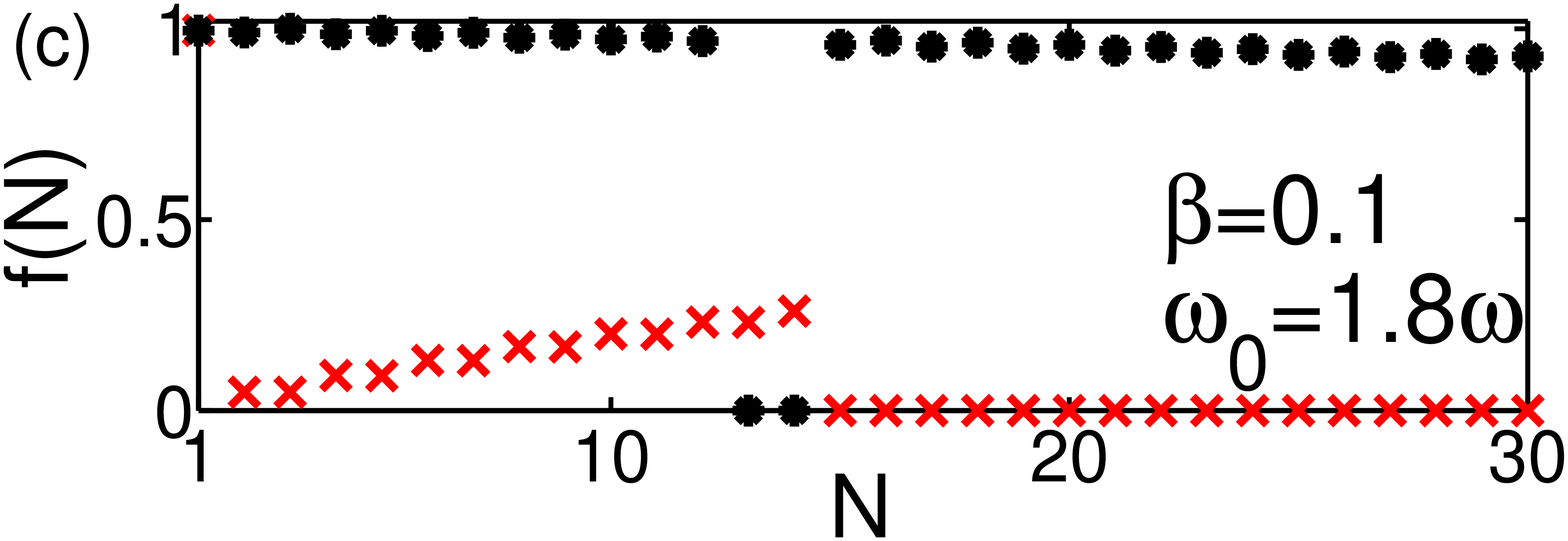}\label{fig:fidelity3}}
\subfigure{\includegraphics[width=4.25cm, height=2.8cm]{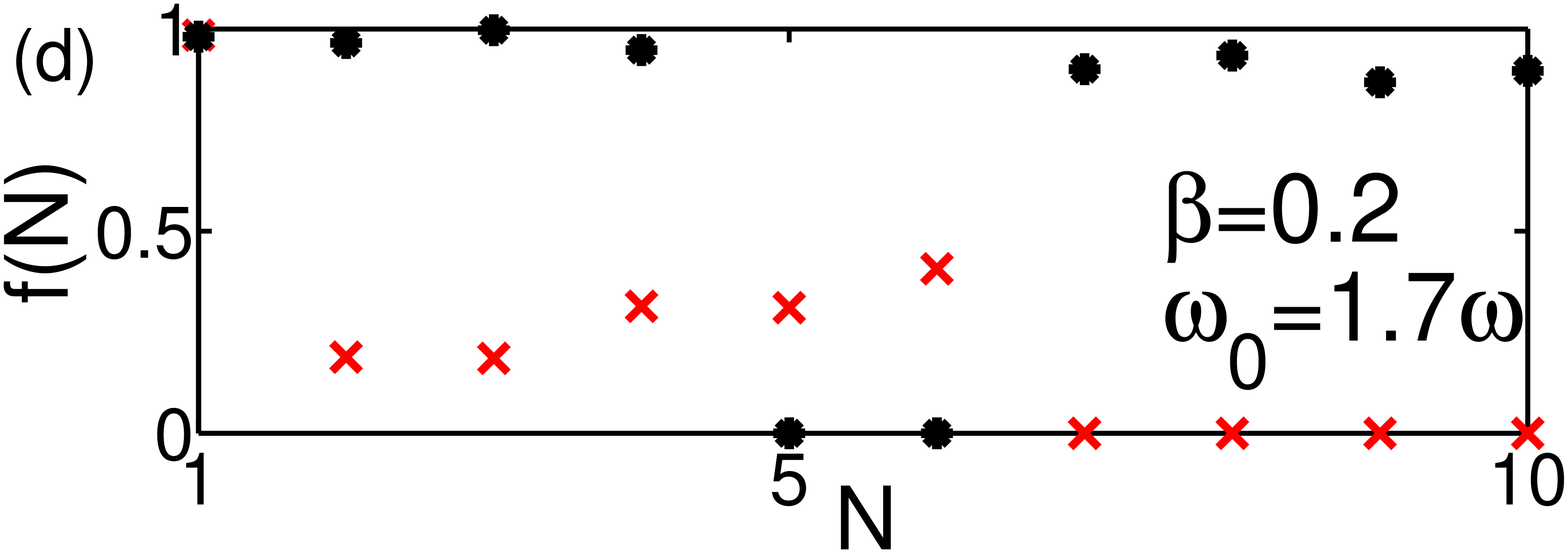}\label{fig:fidelity4}}
\caption{Fidelity of the first $N$ eigenfunctions calculated within the adiabatic approximation (red crosses X) and the Schrieffier-Wolff transformation (black bold dots \Large{$\cdot$}\normalsize). When $\omega_0\ll\omega$, the adiabatic approximation has better fidelity. When $\omega_0\gg\omega$, the adiabatic approximation breaks down.}\label{fig:fidelity}
\end{figure}
Although the eigen-energies predicted by the adiabatic approximation and the SW transformation are the same, we now show that the eigenfunctions calculated within the adiabatic approximation have better fidelity with the exact eigenfunctions than the ones calculated according to the SW transformation. The eigenfunctions calculated within the adiabatic approximation are given in Eq. (\ref{e.adiabatic_eigen}). For the SW transformation, the eigenfunctions are numerically calculated to be $e^{S}\ket{\Psi_{SW}}$, where $\ket{\Psi_{SW}}$ are the eigenfunctions of $H_{SW}$. We now compare these eigenfunctions with the numerically evaluated exact eigenfunctions of $H_{Rabi}$. 

In order to make the comparison, we calculate the fidelity between the approximate eigenfunctions and the exact ones. The fidelity is defined by:
\begin{align}
f=|\scpr{\Psi_{approx}}{\Psi_{exact}}|^2.
\end{align}
For various system parameters, the fidelity for the first few eigenvectors are plotted in Fig. \ref{fig:fidelity}. In Figs. \ref{fig:fidelity1} and \ref{fig:fidelity2}, we choose the system parameters to be such that the qubit frequency is much smaller than the oscillator frequency. In these quasi-degenerate qubit cases, we see that the eigenvectors calculated within the adiabatic approximation have better fidelity than the eigenvectors evaluated according to the SW transformation. On the other hand, as can be seen in Figs. \ref{fig:fidelity3} and \ref{fig:fidelity4}, when the qubit frequency is more than the oscillator frequency, the adiabatic approximation breaks down completely but the SW transformation method still holds. 

Thus, in the quasi-degenerate qubit regime, the adiabatic approximation is better than the SW transformation in approximating the eigen-structure of the Rabi Hamiltonian. In this paper, we only explore the quasi-degenerate qubit regime. For this reason, we carry all our analysis based exclusively on the adiabatic approximation.


\end{document}